\documentclass[aps, prl,10pt,notitlepage,nofootinbib,superscriptaddress,showkeys,twocolumn]{revtex4-1}
\usepackage{amstext,amsmath,amssymb,amsfonts,bbm, amsthm}
\usepackage[latin1]{inputenc}
\usepackage{fancyhdr}
\usepackage{graphicx}
\usepackage{hyperref}

\usepackage{epstopdf}
\usepackage{psfrag}

\usepackage{cases}

\usepackage{subfigure}

\usepackage{tocvsec2}

 \usepackage[usenames,dvipsnames]{pstricks}
 \usepackage{epsfig}
 \usepackage{pst-grad} 
 \usepackage{pst-plot} 

\usepackage{color}

\def\beq{\begin{equation}}
\def\be{\begin{equation}}
\def\ee{\end{equation}}
\def\bes{\begin{eqnarray}}
\def\ees{\end{eqnarray}}

\def\f{\frac}

\begin{document}

\title{\large \bf Entanglement entropy and negative energy in two dimensions}

\author{Eugenio Bianchi}\email{ebianchi@gravity.psu.edu}
\affiliation{Institute for Gravitation and the Cosmos \& Physics Department,
Penn State, University Park, PA 16802, USA}
\author{Matteo Smerlak}\email{msmerlak@perimeterinstitute.ca}
\affiliation{Perimeter Institute for Theoretical Physics, 31 Caroline St.~N., Waterloo ON N2L 2Y5, Canada}

\date{\small\today}

\begin{abstract}\noindent
It is well known that quantum effects can produce negative energy densities, though for limited times. 
Here we show in the context of two-dimensional CFT that such negative energy densities are present in any non-trivial conformal vacuum and can be interpreted in terms of the entanglement structure of such states. We derive an exact identity relating the outgoing energy flux and the entanglement entropy in the in-vacuum. When applied to two-dimensional models of black hole evaporation, this identity implies that unitarity is incompatible with monotonic mass loss.
\end{abstract}
\maketitle

\paragraph{Introduction.}
It has long been known that quantum field theory admits states with local \emph{negative energy densities}, in violation of the positive energy conditions of general relativity. Such negative energies occur for instance in the (static and dynamical) Casimir effect \cite{Davies:1977yv}, with squeezed states of light \cite{Wu:1986jo}, or in particle production in a gravitational field \cite{Hawking:1974sw,Unruh:1977uu}. As pointed out by Ford \cite{Ford:1978fv}, these effects are a potential source of concern, because they apparently lead to a breakdown of the second law of thermodynamics. Fortunately, it turns out that negative energies are necessarily short-lived \cite{Ford:1991cj,Ford:1995hd,Flanagan:2002dq,fewster2005quantum,Blanco:2013lea}: according to ``quantum inequalities'', the product of the absolute value of the negative energy with the characteristic times over which it occurs is bounded from above. Understanding the scope and implications of negative energies is central to gravitational theory, as they are known to lead to classically forbidden gravitational phenomena \cite{Hawking1973}, such as shrinking event horizons and trapped surfaces without singularities.

Here we show that insight into negative energies and constraints thereon can be gained via the concept of \emph{vacuum entanglement entropy}. Entanglement entropy was introduced in quantum field theory by Sorkin \emph{et al.} \cite{Bombelli:1986rw} as a tentative explanation of the origin of the Bekenstein-Hawking black hole entropy. In the context of two-dimensional conformal field theory, the vacuum entanglement entropy of spatial segments---the von Neumann entropy of the partial trace of the vacuum with respect to the exterior of the segment---was computed explicitly in 1994 by Holzhey \emph{et al.} \cite{Holzhey:1994kc}. Since then, entanglement entropy has proved to be an extremely valuable probe of the structure of the vacuum in interacting theories, especially in the vicinity of quantum critical points \cite{Calabrese:2004eu}.

In the context of black hole physics, entanglement entropy was used by Page \cite{Page:1993bv} as a means to investigate the possible outcomes of the Hawking evaporation process \cite{S1975}, where a gravitational collapse results in a quantum energy flux at infinity. Page reasoned as follows. The Hawking radiation and the black hole can be considered as subsystems of an isolated quantum system in a pure state. If the system is finite-dimensional, the entanglement entropy of each subsystem is a bounded non-negative function of the degrees of freedom in each subsystem. Since these degrees of freedom constantly flow between one subsystem (the black hole) and the other subsystem (the radiation), the entanglement entropy of the latter should have two regimes for unitarity to be preserved: a growing phase corresponding to the early Hawking-like stage of evaporation, and a decreasing phase corresponding to the release of information by the ``old'' black hole. The turning point is generally referred to as the ``Page time''.

In this paper we establish a direct connection between negative energies in quantum field theory and unitarity constraints on entanglement entropy. We proceed in three steps. First, we provide a rigorous framework for Page's entropy arguments by giving an explicit, workable definition of the entanglement entropy of conformal vacuum states at future null infinity. Second, we derive an integral identity relating it to the outgoing energy flux, showing in particular that any nontrivial conformal vacuum state must radiate some negative energy. Third, we apply these insights to a class of models of unitary evaporation of ``nonsingular black holes'' discussed in the recent literature \cite{Hayward:2006fn,Frolov:2014wc,Rovelli:2014tm} and comment on their semiclassical mass loss rate.

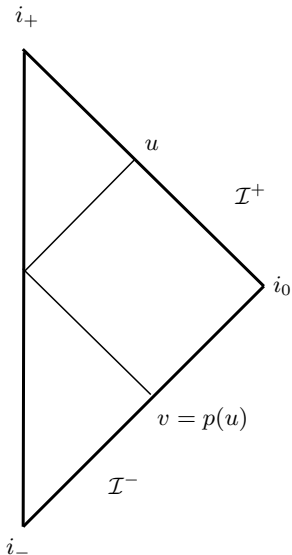
\begin{figure}
\centering
\scalebox{1} 
{
\begin{pspicture}(0,-3.7480469)(4.8729115,3.7480469)
\psline[linewidth=0.04cm](0.41854492,3.0829298)(0.40355495,-3.2303514)
\psline[linewidth=0.04cm](0.40355495,-3.2303514)(3.6035564,-0.030351415)
\psline[linewidth=0.04cm](3.6035564,-0.030351415)(0.39854494,3.1229298)
\usefont{T1}{ptm}{m}{n}
\rput(0.3424707,-3.5253515){$i_-$}
\usefont{T1}{ptm}{m}{n}
\rput(3.8524716,-0.025351416){$i_0$}
\usefont{T1}{ptm}{m}{n}
\rput(1.7524716,-2.6653516){$\mathcal{I}^{-}$}
\usefont{T1}{ptm}{m}{n}
\rput(3.4324718,1.1946483){$\mathcal{I}^{+}$}
\psline[linewidth=0.02cm](1.88,1.6480469)(0.41709054,0.16292973)
\psline[linewidth=0.02cm](0.39709052,0.20292962)(2.1,-1.4719532)
\usefont{T1}{ptm}{m}{n}
\rput(2.7924707,-1.8053515){$v=p(u)$}
\usefont{T1}{ptm}{m}{n}
\rput(0.4624707,3.5546484){$i_+$}
\usefont{T1}{ptm}{m}{n}
\rput(2.1124709,1.8546485){$u$}
\end{pspicture} 
}
\caption{Conformal diagram of an asymptotically flat half-plane with a reflecting boundary, defining a canonical mapping $v=p(u)$ between future null infinity $\mathcal{I}^{+}$ and past null infinity $\mathcal{I}^{-}$.}
 \label{confdiag}
        \end{figure}

\paragraph{Conformal vacuum.}  We consider an asymptotically-flat two-dimensional spacetime $\mathcal{M}$ with metric $ds^2=-\Omega^2(u,v)dudv$ and double-null coordinates $(u,v)$ satisfying the inequality $v\geq p(u)$ for some continuous one-to-one and onto function $p(u)$. We assume that $\lim_{u\to -\infty} \dot{p}(u)=1$, where $\dot{p}\equiv dp/du$, and that the coordinates $u$ and $v$ induce affine parameters on future and past null infinity, denoted $\mathcal{I}^{+}$ and $\mathcal{I}^{-}$ respectively\footnote{We assume $\lim_{u\to -\infty} \Omega^2(u,v)=\lim_{v\to +\infty}\Omega^2(u,v) =1$ so that, asymptotically, the coordinates $v$ and $u$ match the proper time along the orbit of a future-pointing asymptotic time-like Killing vector of $\mathcal{M}$.}. The spacetime $\mathcal{M}$ can be thought of alternatively as a ``moving mirror'' in flat space \cite{Davies:1977yv,Walker:1985dwa,Carlitz:1987hd}, or as the time-radius plane of a spherically symmetric four-dimensional spacetime (the boundary $v=p(u)$ representing the center $r=0$). 

We consider a quantum field propagating on $\mathcal{M}$, with reflecting boundary condition along $v=p(u)$. The asymptotically-flat structure of  $\mathcal{M}$ provides a natural notion of vacuum state (and of particles) at past null infinity $\mathcal{I}^-$ and at future null infinity $\mathcal{I}^+$. We call these two (Heisenberg) states the in-vacuum $|0_-\rangle$ and the out-vacuum $|0_+\rangle$, respectively. For a conformal field theory with central charge $c$, these two states are related by a unitary transformation $U[p]$ that acts on primary fields as 
\begin{equation}
U[p] \Phi(u)U[p]^\dagger=\dot{p}(u)^h\,\Phi(p(u))\,.
\label{eq:unitary}
\end{equation}
Importantly, while the renormalized energy-momentum tensor $T_{\mu\nu}$ of the out-vacuum $|0_+\rangle$ vanishes at $\mathcal{I}^+$, for the in-vacuum $|0_-\rangle$ it has a non-trivial component, $F=T_{\mu\nu}(d/du)^{\mu}(d/du)^{\nu}$, given by the Davies-Fulling-Unruh formula \cite{Davies:1976uk,Birrell1982}
 \be
F_{|0_-\rangle}(u)=-\f{c}{24\pi}\left(\f{\dddot{p}(u)}{\dot{p}(u)}-\f{3}{2}\f{\ddot{p}(u)^{2}}{\dot{p}(u)^{2}}\right)\,.
\label{flux}
\ee
Thus, a conformal field prepared in the vacuum $|0_-\rangle$ in the past and let to evolve in a spacetime $\mathcal{M}$ with a non-trivial $p(u)$, results in outgoing radiation that reaches future null infinity; formula (\ref{flux}) expresses the energy flux of this radiation.

\paragraph{Renormalized entanglement entropy.} Consider an asymptotic observer at $\mathcal{I}^+$ measuring the field in a state $|s\rangle$ for a finite time $A=[u_0,u\,]$. The results of all her measurements are coded in the reduced density matrix $\rho_A=\text{Tr}_{\bar{A}}(|s\rangle\langle s|)$ where modes in the complement $\bar{A}$ have been traced away. The density matrix $\rho_A$ is mixed because of quantum correlations between $A$ and its complement. A measure of such correlations between two disjoint intervals $A$ and $B$ is provided by the mutual information $I(A,B)$, i.e. the relative entropy of the density matrix $\rho_{A\cup B}$ with respect to the product matrix  $\rho_{A}\otimes \rho_{B}$ \cite{Casini:2008wt}, 
\begin{equation}
I_{|s\rangle}(A,B) \equiv\text{Tr}\big(\rho_{A\cup B}\log \rho_{A\cup B}-\rho_{A\cup B}\log\rho_{A}\!\otimes \rho_{B}\big).
\end{equation}
This quantity is finite as long as the intervals $A$ and $B$ are disjoint \cite{araki1975relative}. 

Consider now a broadening $A\cup \Delta$ of the region $A$, with $\Delta\equiv\Delta_0\cup\Delta_1=[u_0-\delta u_0,u_0)\cup (u,u+\delta u]$ and $\delta u_0>0$, $\delta u>0$. We define the \emph{(covariantly regularized) entanglement entropy} of $A$ as one half of the mutual information between $A$ and the complement of its broadening $B=\;\overline{\!A\cup\Delta\!}\;$, i.e.
\begin{equation}
S_{|0_+\rangle}^{\Delta}(A)\equiv\frac{1}{2}I_{|s\rangle}(A,\,\overline{\!A\cup\Delta\!}\;)\,.
\end{equation}
The region $\Delta$ that separates $A$ from $B$ plays the role of covariant regulator that can be removed by  taking the limits $\delta u_0\to 0$, $\delta u\to 0$. 

The entanglement entropy $S_{|0_+\rangle}^{\Delta}(A)$ of the out-vacuum state $|0_+\rangle$ can be easily computed using three ingredients: $(i)$ the expression given by Holzhey \emph{et al.} \cite{Holzhey:1994kc} for the entanglement entropy $S_\epsilon(A)=-\text{Tr}(\rho_A \log \rho_A)_{\epsilon}$ of a segment $A$ in the presence of a UV cut-off $\epsilon$,
\begin{equation}
S^{\epsilon}_{|0_+\rangle}(A)=\frac{c}{12}\log \;\frac{(u-u_0)^2}{\epsilon^2},
\end{equation}
$(ii)$ the relation between mutual information and the cut-offed entropy \cite{Casini:2008wt}
\begin{equation}
I_{|s\rangle}(A,\,\overline{\!A\cup\Delta\!}\;)=S_{|s\rangle}^{\epsilon}(A)+S_{|s\rangle}^{\epsilon}(A\cup \Delta)-S_{|s\rangle}^{\epsilon}\Delta,
\end{equation}
and $(iii)$ the fact that in the limit $\delta u_0\to 0$, $\delta u\to 0$ the entropy of the union of two disjoint intervals reduces to the sum of the entropies of each interval, i.e. $S_\epsilon(\Delta_0\cup\Delta_1)\to S_\epsilon(\Delta_0)+S_\epsilon(\Delta_1)$ \cite{Casini:2008wt}. The resulting expression is independent of the cut-off $\epsilon$ and reads
\begin{equation}
S^{\Delta}_{|0_+\rangle}(A)=\frac{c}{12}\log\frac{(u-u_0)^2}{\delta u\;\delta u_0}\,.
\label{S0}
\end{equation}

Now, the transformation $U[p]$ defined in Eq.~(\ref{eq:unitary}) acts unitarily on states and the regularized entanglement entropy transforms covariantly as
\begin{equation}
S_{U[p]|0_+\rangle}^{\Delta}(A)\,=S_{|0_+\rangle}^{p^*\!\Delta}(p^*\!A)
\label{SU}
\end{equation}
where $p^*\!A$ is the pull-back of the interval $A$ through the the diffeomorphism $u\mapsto p(u)$. Using formulae (\ref{S0}) and (\ref{SU}) we find that the regularized entanglement entropy of the conformal vacuum $|0_-\rangle=U[p]|0_+\rangle$ is given by
\begin{equation}
S^{\Delta}_{|0_-\rangle}(A)= \frac{c}{12}\log\frac{(p(u)-p(u_0))^2}{\dot{p}(u)\,\dot{p}(u_0)\,\delta u\,\delta u_0}\,.
\label{S1}
\end{equation}

We are now ready to define the \emph{renormalized entanglement entropy} $S(u)$ of the radiation in the state $|0_-\rangle$ up to a time $u$ by subtracting the out-vacuum contribution and taking the limit $u_0\to-\infty$,
\begin{equation}
S(u)\equiv\lim_{u_0\to -\infty}\; S^{\Delta}_{|0_-\rangle}([u_0,u])-S^{\Delta}_{|0_+\rangle}([u_0,u]).
\end{equation}
From (\ref{S0}), (\ref{S1}) and the assumption that $\dot{p}(u)\to 1$ as $u\to-\infty$, we find that the renormalized entanglement entropy of the radiation up to a time $u$ is simply given by the formula
\begin{equation}
S(u)=-\frac{c}{12}\log\dot{p}(u)\,.
\label{entropy}
\end{equation}
Further insight into formula \eqref{entropy} can be gained by introducing the ``peeling function'' $\kappa\equiv-\ddot{p}/\dot{p}$ familiar from Hawking radiation theory\footnote{When the adiabaticity condition $\vert \dot{\kappa}/\kappa^{2}\vert\ll1$ is satisfied, $\kappa$ can be interpreted as ($2\pi$ times) the instantaneous temperature of the outgoing flux \cite{Barcelo:2011cf}.} \cite{Barcelo:2011cf}, this gives
\be\label{peeling2}
S(u)=\f{c}{12}\int_{-\infty}^{\,u}\kappa(u')du'\,.
\ee
Eq. \eqref{peeling2} provides us with an intuitive interpretation of entanglement entropy production at $\mathcal{I}^{+}$: $S(u)$ grows when the outgoing null geodesics are \emph{peeled} ($\kappa(u)>0$), and decreases when they are \emph{squeezed} ($\kappa(u)<0$). The curve $S(u)$ is often referred to as the ``Page curve'' \cite{Page:1993bv} in the black hole literature.

\paragraph{Energy-entropy relation at future null infinity.} We now investigate the relationship between entanglement entropy and energy flux. Combining \eqref{entropy} and \eqref{flux}, we obtain
\be\label{riccati}
F(u)=\f{1}{2\pi}\left(\f{6}{c}\dot{S}(u)^{2}+\ddot{S}(u)\right).
\ee
This identity implies that the outgoing flux $F$ is completely determined by the structure of entanglement at future null infinity. Reciprocally, since \eqref{riccati} is a second-order differential equation in $S(u)$, the entanglement entropy is completely determined by the flux and the values of $S(u)$ and $\dot{S}(u)$ at one point of $\mathcal{I}^{+}$. The latter are not free: by construction we have $S(-\infty)=0$, and for the total outgoing energy to be finite, $\int_{\mathcal{I}^{+}}du\,F(u)<\infty$, equation \eqref{riccati} requires that 
\be\label{limits}
\dot{S}(u)\to0\quad\textrm{as}\quad u\to\pm\infty.
\ee
Note that $S(+\infty)$ and $S(-\infty)$ can be different: in the case of an asymptotically inertial ``moving mirror'' with different initial and final velocities \cite{Walker:1985dwa}, one gets 
$S(+\infty)=-c\,\eta/6$, where $\eta$ is the rapidity of the final rest frame relative to the initial rest frame. 

\begin{figure*}[t!]
\psfrag{bb}{\hspace{-1.7em}$\frac{S(u)}{S_{\text{max}}}$}
\psfrag{aa}{\hspace{-.8em}$u/\tau$}
\psfrag{dd}{\hspace{-1.7em}$\frac{F(u)}{F_H}$}
\psfrag{cc}{\hspace{-.8em}$u/\tau$}
\includegraphics[height = 14em]{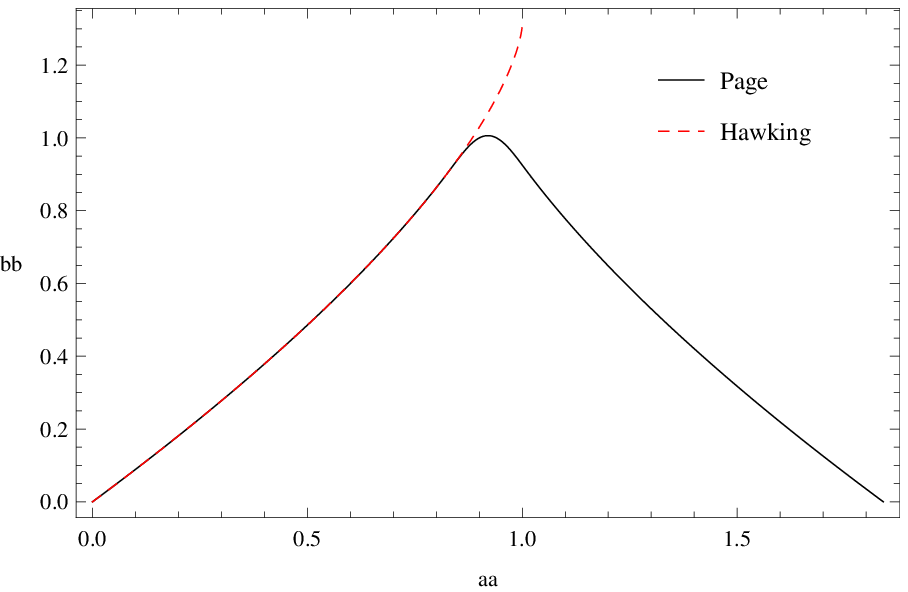}\hspace{7em}
\includegraphics[height = 14em]{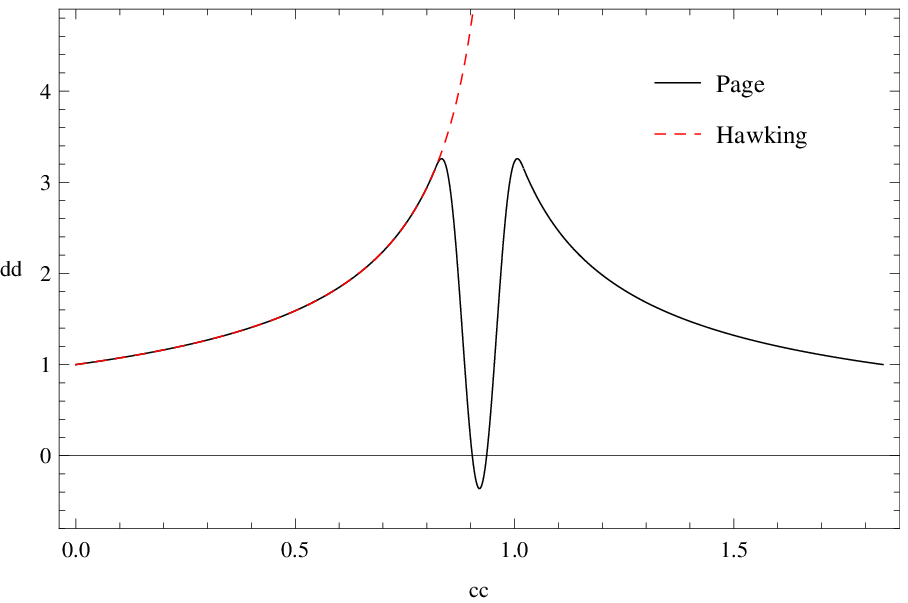}
\caption{Left: Entanglement entropy in black hole evaporation, as posited by Page \cite{Page:1993bv}. Right: The corresponding flux function (normalized to the Hawking flux $F_{H} \sim\hbar/M_{\textrm{ADM}}^{2}$), as derived from \eqref{riccati}. (The Hawking thermal entropy and mass law (dashed line) are for reference, and $\tau\sim M_{\textrm{ADM}}^{3}/\hbar$ denotes the Hawking evaporation time.)}
\label{page}
\end{figure*}
 
\paragraph{Mapping to a scattering problem.}
The scope of this energy-entropy relation is most clearly revealed by interpreting Eq.~\eqref{riccati} as a differential equation for the entropy $S(u)$, with the flux $F(u)$ as given. To solve Eq.~\eqref{riccati} we note that it has the form of a Riccati equation and it can be put in linear form introducing the auxiliary functions $\psi$ and $V$ defined as 
\be
\dot{S}(u)\equiv \f{c}{6}\,\f{\dot{\psi}(u)}{\psi(u)}\quad\textrm{and}\quad
V(u)\equiv\f{12\pi}{c}\,F(u).
\ee
Standard manipulations then reveal that \eqref{riccati} reduces to 
\be\label{schrodinger}
-\ddot{\psi}(u)+V(u)\,\psi(u)=0.
\ee
This equation can be interpreted as the Schr\"odinger equation for a particle scattering on the potential $V$ with zero energy. Since $\psi(u_{1})/\psi(u_{0})=\exp[6(S(u_{1})-S(u_{0}))/c]$, unitarity requires that $\psi$ has constant sign and finite limits $\psi(\pm\infty)$. From the perspective of quantum scattering theory, this corresponds to a \emph{zero-energy resonance} (or \emph{half-bound state}) \cite{Senn:1988gj}. The existence of such states\footnote{Low-energy resonances were first hinted at experimentally by Ramsauer and Townsend in the form of anomalously high transmission coefficients at low energy \cite{RamsauerTownsend:1921}.} places sharp constraints on the potential $V$. In particular, one can show by writing \eqref{schrodinger} in Volterra form 
\be
\psi(u)=\psi(-\infty)+\int_{-\infty}^{u}du'\, (u-u')\,V(u')\,\psi(u')
\ee
and solving it by the Neumann series method that such ``exceptional potentials'' are unstable: a generic perturbation of $V$ will \emph{not} admit a zero-energy resonance. An obvious necessary condition on $V$ for it to admit such states is that $V$ is not positive definite: by integration of \eqref{schrodinger} and using \eqref{limits}, we have
\be
\int_{-\infty}^{\infty}du\, V(u)\,\psi(u)=0.
\ee
Thus, the entanglement entropy at $\mathcal{I}^{+}$ can be expressed in terms of the outgoing flux as the series
\begin{widetext}
\be
S(u)=\f{c}{6}\log \left[1+\sum_{m=1}^{\infty}\left(\f{12\pi}{c}\right)^{m}\int_{-\infty}^{u}du_{1}\int_{-\infty}^{u_1}du_{2}\cdots\int_{-\infty}^{u_{m-1}}du_{m}\prod_{i=1}^{m}(u_{i-1}-u_{i})\,F(u_{i}) \right],
\ee
\end{widetext}
where $u_{0}\equiv u$. Furthermore,
\be\label{mainresult}
\int_{\mathcal{I}^{+}}du\, F(u)\, e^{6S(u)/c}=0.
\ee
Eq. \eqref{mainresult} is the main result of this paper. It shows that \emph{any geometry affecting the entanglement entropy of the conformal vacuum at $\mathcal{I}^{+}$} (e.g. any asymptotically inertial ``moving mirror'' trajectory) \emph{must radiate some amount of negative energy}.  In other words, transient violations of the null energy conditions are not features of peculiar phenomena such as the Casimir and Hawking effects---they are a property of any nontrivial conformal vacuum state. 

More detailed information about the relation between the negative energy flux and the entanglement entropy can be obtained directly from \eqref{schrodinger}. It is immediate to show that, when the flux is positive, the entanglement entropy is either strictly increasing or strictly decreasing. On the other hand, when the flux is negative, $\psi(u)$ is an oscillating function and $\dot{S}(u)$ can change sign. Therefore a Page curve as in Fig.~\ref{page} requires at least a phase with a negative energy flux.\footnote{In fact, when $u(u)$ can be extended to a diffeomorphism of the real projective line, i.e. when $u\mapsto1/u(1/u)$ is smooth at $u=0$, a recent theorem of Ghys on the zeros of Schwarzian derivatives \cite{Ovsienko:2005tq} implies that $F$ must change sign at least \emph{three} times before reaching $\lim_{u\to\infty}F(u)=0$.} 
 Moreover the requirement that the function $\exp[6\,S(u)/c]$ remains positive (i.e. that the renormalized entanglement entropy of the conformal vacuum is bounded from below and never reaches $-\infty$) puts an upper bound on the allowed duration of the oscillating phase where a negative energy flux is emitted. These results are fully compatible with quantum energy inequalities \cite{Ford:1991cj,Ford:1995hd,Flanagan:2002dq,fewster2005quantum,Blanco:2013lea}.

\paragraph{Application to unitary black hole evaporation.}

Let us now consider the implications of our results for the ``information loss problem'' \cite{Hawking:1976wi} in black hole physics. Suppose that a spherically symmetric collapsing matter distribution forms a black hole with ADM mass $M_{\textrm{ADM}}$. If one neglects backscattering, the $s$-wave sector of the Hawking radiation (which is expected to carry the bulk of the radiated energy) can be described by a two-dimensional massless conformal field theory. The results in the previous sections then imply that, for black hole evaporation to be consistent with a classical spacetime with the causal structure of Minkowski space \cite{Hayward:2006fn,Frolov:2014wc,Rovelli:2014tm}, the retarded mass of the hole 
\be
M(u)\equiv M_{\textrm{ADM}}-\int_{\mathcal{I}^{+}}du\, F(u)
\ee
cannot be monotonically decreasing. To illustrate this conclusion, we plot in Fig.~\ref{page} the flux $F(u)$ implied by a symmetric entropy curve $S(u)$, as posited by Page \cite{Page:1993bv}. When reaching the Page time, the flux suddenly (and shortly) becomes negative, leading to a transient increase of the retarded mass $M(u)$ of the hole. Pictorially, the black hole ``gasps'' before dying a unitary death.

\paragraph{Conclusion.}

We have studied the relationship between energy flux and entanglement entropy at future null infinity in two-dimensional conformal vacuum states. By means of a Schr\"odinger-like differential equation relating them, we have obtained strong constraints on the outgoing energy flux. In particular, we have showed that any nontrivial conformal vacuum state must contain some negative energy at future null infinity. Reciprocally, we have showed that the Page curve for the entanglement entropy is completely determined by the flux function. Our results apply to the Hawking phenomenon---where they suggest that unitarity implies non-monotonic evaporation, but also to dynamical Casimir (moving mirrors) effects, and more generally to any ``squeezed'' state of a two-dimensional massless field \cite{BSsqueezed}. 

\paragraph{Acknowledgements.} We thank Ted Jacobson, Abhay Ashtekar and Carlo Rovelli for useful discussions, and Etienne Ghys for kindly answering our questions about Schwarzian derivatives. Research at the Perimeter Institute is supported in part by the Government of Canada through Industry Canada and by the Province of Ontario through the Ministry of Research and Innovation.


\providecommand{\href}[2]{#2}\begingroup\raggedright\endgroup

\end{document}